**Field of View Based Optimization of Aspheric Designed Geometric-Phase Doublet Lenses**

Kathryn J. Hornburg[1,2,*], Xiao Xiang[1,3], Michael W. Kudenov[1], and Michael J. Escuti[1,4]

[1] *Department of Electrical and Computer Engineering, North Carolina State University, Raleigh, NC, United States*
[2] *Current Affiliation: Duke Center for In Vivo Microscopy, Department of Radiology, Duke University Medical Center, Durham, NC, United States*
[3] *Current Affiliation: Apple Inc., Cupertino, California, United States*
[4] *Current Affiliation: Meta Platforms Inc. (dba Meta), Menlo Park, California, United States*

*Correspondence:
Kathryn J. Hornburg
kathryn.hornburg@duke.edu

Manuscript Length: 5381 words

**ABSTRACT**

Here we study aspheric phase pattern geometric-phase lenses through the design of doublet lenses in Zemax Optic Studio and in MATLAB using a simple ray tracing code. The solutions developed use variable spacer distances and are solved monochromatically (at 633 nm) and polychromatically (at 450, 532, and 633 nm). After selecting a Zemax generated polychromatically optimized doublet solution, we fabricated the lens utilizing liquid crystal films and direct writing. We characterized the liquid crystal alignment quality, efficiencies, and spot performance of this lens in comparison to a standard spherical singlet of the same back focal length and diameter (24.5 mm diameter and 40 mm back focal length at 633 nm). With this aspheric lens profile, we realize the development of novel focusing phase surfaces.

# INTRODUCTION

Geometric-phase lenses provide a substantial reduction in the thickness for similar on-axis performance of traditionally manufactured lenses[1, 2], providing a solution for biomedical optical systems[3] and heads-up display technology [4]. Implementation of traditional geometric-phase lenses into imaging system is hampered by an inherent limitation of the spherical phase profile [5, 6].Illustrated in **Figure 1A**, we see an ideal spot with on-axis lighting since the focal length $f_0$ and wavelength utilized match those selected in the spherical phase profile of the hologram. As incoming light becomes less on-axis, see **Figure 1B**, the resulting focused spot profiles contain more aberration, such as coma, since the spherical phase profile does not take in account the off-axis lighting condition. This limitation is most apparent in cases where angular information is needed to render a scene, such as plenoptic imaging [7], since the data off axis needs to be maintained with the same resolution as on axis to obtain a sharp combination of images. A common way to reduce off-axis aberration is to generate aspheric phase profiles and spaced doublets [8]. In this study, we develop various doublet schemes utilizing MATLAB and Zemax OpticStudio 17.5 generated phase surfaces, allowing for potentially aspheric solutions. We compare the solutions found for both methods then manufacture a standard spherical phase singlet and optimal doublet design for analysis. We compare the Zemax OpticStudio 17.5 solutions to that of the original optimization results from Zemax OpticStudio 16.5 presented in two conference proceedings[5, 6].

# BACKGROUND

To generate the beam shaping designs of these aspheric lenses, we utilize geometric phase over propagation phase[9, 10]. Meaning, rather than spatially vary optical path length via refractive index or physical thickness, we spatially vary the polarization state of the light[11, 12]. This gives rise to a beam shaping effect, called the Pancharatnam-Berry effect, with the accumulated "memory" generating a phase offset in a thin-form factor. In the case of this work, we utilize liquid crystals, where the varying alignment of the birefringent media $\Phi(x,y)$ make for a polarization change across the lens surface.

The geometric phase contribution is described by:

$$\delta(\boldsymbol{r}, f_o, \lambda_o) = \pm 2\Phi(x, y) \ ,$$

with the standard spherical phase profile of the geometric phase lens described as [1]:

$$\delta_\pm(\boldsymbol{r}, f_o, \lambda_o) = \mp \frac{2\pi}{\lambda_o}\left(\sqrt{r^2 + f_o^2} - f_o\right) \ .$$

A similar technique can be used for metasurface based elements [13]; a class of nanoscale structured light bending devices where optical resonators generate the spatially varying pattern.

The phase pattern generates three optical channel focusing at $\pm f$ and a leakage channel due to the nature of continuous phase variation in geometric phase diffraction elements [14, 15].The efficiency of those channels is given by:

$$\eta_o = (\cos \Gamma/2)^2$$

and

$$\eta_{\pm 1} = \tfrac{1}{2}(1 \mp S_3')(\sin \Gamma /2)^2,$$

where $\Gamma$ is the optical retardation specified by $2\pi\Delta nd/\lambda$ and $S_3'$ is the circular Stokes term $S_3$ normalized to the total intensity of beam channel $S_0$. When half-wave conditions are maintained with continuous phase variation, such as that possible by liquid crystal films, the leakage term approaches 0%. We can select the desired optical path for focusing by using appropriately polarized light [15]. Inherent to having thin lenses with this phase profile, there is introduced aberrations as one utilizes off-axis illumination [16]. Prior metalens studies indicate aplantic and aspheric surfaces can be used to improve off-axis imaging performance [17, 18].

**STUDY LAYOUT**

In this study, we develop models of doublet geometric-phase lenses. As an aside, previously we uncovered that singlet aspheric solutions only re-balance focus [6] and therefore do not include the discussion of those lenses here. Our model doublet lenses are 24.5 mm diameter with 40 mm back focal lengths at 633 nm. We represent the lens systems in Zemax OpticStudio using the Zernike fringe description of a holographic phase surface and more simply in MATLAB by generating a basic ray tracing code. Schematics of both lens systems are presented in **Figure 2** where **D** indicates our diameter and **BFL** the back focal length. By comparing both tracing systems (Zemax and MATLAB), we look for the optimal design with different phase expansions and determine if a similar trend can be verified in both models. The end caps (0.21 mm) and spacers (variable component, **L**) are modeled by N-BK7. In the MATLAB model, end caps are not included. There is minimal impact on the final spot size by not including endcaps.

**KEY METRICS**

The spot size of the focused beam is utilized as our optimization metric plus the inclusion of standard deviation across colors or angles. Zemax provides a rms spot size radius calculation based on the root mean square distance of a point in the ray fan and a desired reference location, such as the chief ray, for each given ray fan. Thus, it provides more information of the ray spread, which the aberration concerns relate to, than the geometric spot size calculation. In MATLAB, we develop a similar criterion to generate our own rms spot size radius calculation matching that above. To compare the different phase profiles generated, we illustrate the phase and period of individual selected lenses. Piston is not seen as an aberration since it is a phase offset of the whole of the optic and thus also adds complexity in comparison when we have disparate values of piston for each lens in our system. Thus, we remove piston from our phase by setting the phase equal to 0 at the center of the lens, $\delta_\pm (0) = 0$. The phase in our graphs is presented as the absolute value of the relative phase shift of each lens.

To determine the back focal length of our extension wavelengths (450 and 532 nm), we generally assume the same write to replay relationship as other studies [19, 20], but allow adjustments to the back focal length for the Zemax monochromatic simulation since the equation's approximation conditions of small incident angles might not hold for large spacer distances. The relationship generally assumed is:

$$f(\lambda) = \frac{f_o \lambda_o}{\lambda}.$$

For the Zemax polychromatic and MATLAB simulations, we fix the back focal lengths to 48 and 57 mm, for 450 and 532 nm respectively, which represent approximately the back focal length response of the Zemax monochromatic 0 to 10 mm glass spacers simulations.

*ZEMAX Model*

In Zemax, we develop the asphere holographic GP lens models using field rays of 0°, ± 3° and ± 7° along X- and Y- axes. The holographic surface uses Zernike fringe phase coefficients which is a method to radially represent wavefront [21, 22]. After early analysis of the Zernike fringe terms for both singlet and doublet cases, it was found that the phase terms corresponding to power ($Z_3$), primary spherical ($Z_8$), and secondary spherical ($Z_{15}$) give the best response overall and are the tuned Zemax parameters for this work. All other Zernike fringe parameters were set to zero. Thus, the phase equation in Zemax we work to minimize error on is:

$$\delta_\pm(r) = \mp 2\pi(Z_3(-1 + 2r^2) + Z_8(1 - 6r^2 + 6r^4) + Z_{15}(-1 + 12r^2 - 30r^4 + 20r^6)).$$

We present the Zernike Fringe terms we discover through optimization as $\mathbf{Z_L}$= [$Z_3$, $Z_8$, $Z_{15}$], where **L** specifies the lens number in the possible system.

Initial studies were done utilizing Zemax OpticStudio 16.5 [5, 6]. In this work, studies were primarily completed utilizing Zemax OpticStudio 17.5 which has an updated optimization function that considers x and y slices of the spot radius separately [23]. Although, in both cases gaussian quadrature sampling with 20 rings and 12 arms was utilized for optimizing the rms spot size.

*MATLAB Model*

To further explore the optimal design with different phase expansions, we develop a model in MATLAB to simulate a variety of diffractive lenses. In this customized model, the ray tracing is based on the generalized law of refraction:

$$n_t \sin\theta_t = \frac{\lambda}{\Lambda(r)} + n_i \sin\theta_i,$$

with incidence *i* and exit *t* medium respectively. $\Lambda(r)$ indicates the period of a given lens surface at a distinct point on the radius of the lens. After the exit angle $\theta_t$ of the lens calculated, the ray is simply transferred through the same medium using cartesian advancements of the exit angle. From testing multiple phase expansions for the same simulation setup conditions, it was determined that a simple polynomial phase expansion following:

$$\delta_\pm(r) = \mp 2\pi \sum_{i=1}^{N} C_i r^{i+1},$$

produces optimal results with our optimization terms as $C_i$ and letting i=1… 5. In the similar way as the Zernike generated surfaces, the final coefficient terms are presented as $\mathbf{C_L}$= [$C_1$, $C_2$, $C_3$, $C_4$, $C_5$], where **L** specifies the lens number in the possible system. The final lens phase optimization was performed with the built-in local and global algorithms in MATLAB.

**RESULTS**

*Monochromatic Doublet Optimization*

For doublet monochromatic optimization, we begin by utilizing 633 nm as the optimization wavelength. All results are shown in **Figure 3**.

Discussing first the process for both models, we generate an optimization procedure for 0.25 and 70 mm for Zemax simulations and 0.5 to 70 mm for MATLAB in non-uniform steps. There are two regimes for the optimization. One, from 0 to 10 mm spacer distance and two, from 10 to 70 mm spacer distance. For the small spacer distance, the trend of the MATLAB corresponds highly with the Zemax result for the red wavelength response, but deviates for blue and green. For the large spacer distances, we see an offset in the red results, but a same general trend, while the deviation of blue and green continue as from the smaller spacer distances. We sparsely resimulated the monochromatic data using the 17.5 optimizer to determine if the updated optimizer provides a better solution for the extension wavelengths. In comparison to the Zemax OpticStudio 16.5 solutions, we saw that the solutions found from the updated 17.5 optimizer match that of the prior solution, solid versus diamond markings in **Figure 3**, therefore we continue to use the Zemax 16.5 solutions here for full analysis. This logically follows because the red monochromatic case we are studying is already minimal much in same way as the MATLAB solution indicating the condition, we are optimizing for is already minimal with the initial optimizer formula.

Selecting the 1 mm spacer distance solution of both models, the optimal coefficients for the monochromatically minimized doublet phase profiles are:

$$Z_1 = [-21.46, 2.77 \times 10^{-2}, 2.84 \times 10^{-6}],$$
$$Z_2 = [12.51, -3.16 \times 10^{-2}, 2.59 \times 10^{-6}]$$

and

$$C_1 = [9.95, -0.187, -4.58 \times 10^{-2}, -1.26 \times 10^{-2}, 3.95 \times 10^{-4}],$$
$$C_2 = [9.85, 0.059, 7.45 \times 10^{-2}, 1.04 \times 10^{-2}, -4.20 \times 10^{-4}].$$

We present the period and absolute phase of these solutions in **Figure 4**. The peaking in the period profile curves indicate that there are regions of both convergence and divergence on the same lens. The average rms spot size radii for the models at 1 mm glass separation distance and the three test wavelengths are for Zemax, 72.1 ± 37.9 µm (red: 61.2 ± 19.0 µm) and with MATLAB 192 ± 16.4 µm (red: 79.1 ± 5.56 µm). The minimum periods for the Zemax generated aspheric model are 2.67 µm and 1.45 µm for Lens 1 and 2 respectively, which compares to a MATLAB minimized solution with minimal periods of 1.07 µm for Lens 1 and 0.78 µm for Lens 2. The MATLAB generated lenses would be more complex manufacture due to the smaller minimal periods.

*Polychromatic Doublet Optimization*

Since the monochromatic result only produced a cohesive trend for the red wavelength, we move to optimizing polychromatically by adding the extension wavelength into the simulation. The polychromatic optimized models use 0.5 to 70 mm glass spacer distances in non-uniform steps. The resulting spot sizes comparing between Zemax versions are shown in **Figure 5**. We see a drastic improvement in the variation of spot radii with the updated Zemax 17.5 optimizer for the large spacer distances and therefore turn to MATLAB to determine the similarity and differences with Zemax 17.5.

The resulting angle averaged rms spot size radii are shown in **Figure 6** for the polychromatically optimized Zemax 17.5 and MATLAB models. The spot size radii are more similar across wavelength with this minimization method. In **Figure 6A**, we see that the MATLAB and Zemax generated results agree for both thin and thick lens systems. Across models, the average difference of the red spot sizes is 0.57 µm. Investigating the green and blue wavelength results, **Figure 6B**, a similar agreement occurs across both models. The average differences in spot sizes are 3.8 µm for green and 8.0 µm for blue, thus showing agreement in the optimal solution. The minimal rms spot radii occur at 70 mm separation for both systems. The average rms spot size radii with the 70 mm spacer are 16.7 ± 3.3 µm (red: 16.1 ± 2.6 µm) and 17.2 ± 6.2 µm (red: 19.8 ± 4.3 µm) for Zemax and MATLAB respectively.

Selecting the 1 mm spacer distance solution of both models since we desire a balance of lens thickness and performance, the optimal coefficients for the polychromatically minimized doublet phase profiles are:

$$Z_1 = [-22.30, 2.16 \times 10^{-2}, 2.31 \times 10^{-6}],$$
$$Z_2 = [13.36, -2.44 \times 10^{-2}, 6.34 \times 10^{-7}]$$

and

$$C_1 = [33.85, -4.29 \times 10^{-2}, -0.102, -2.17 \times 10^{-4}, -7.88 \times 10^{-5}],$$
$$C_2 = [-15.23, 4.49 \times 10^{-2}, 0.108, 9.82 \times 10^{-4}, -1.99 \times 10^{-8}].$$

We present the period and absolute phase of these solutions in **Figure 7**. Like before, the peaking in the period profile curves indicate that there are regions of both convergence and divergence on the same lens, but the first Zemax generated lens this time shows no phase reversal, just a minimal pitch region that is now towards the inside of the lens rather than the edge. The minimum periods for the Zemax generated aspheric model are 2.24 µm and 2.54 µm for Lens 1 and 2 respectively, which compares to a MATLAB minimized solution with minimal periods of 3.16 µm for Lens 1 and 1.81 µm for Lens 2. The complexity of both solutions to manufacture is similar. The average rms spot size radii for the models at 1 mm glass separation distance and the three test wavelengths are for Zemax 68.4 ± 17.34 µm (red: 71.5 ± 8.67µm) and with MATLAB 62.5 ± 36.5 µm (red: 73.7 ± 36.6 µm).

*Fabrication of Test Lenses*

We select the polychromatic Zemax result for manufacture because of the slightly reduced spot size in comparison to the monochromatic results and less expected variation of spot size over MATLAB simulations. To verify the optimization, we manufactured two geometric-phase lens with the different aspherical phase profiles, corresponding to the 1 mm spacer polychromatic

Zemax optimization, and assembled them into the doublet configuration in order to evaluate normal and oblique incidence performance. A spherical phase singlet with the same BFL was generated for comparison. The fabricated GPL layouts are shown in **Figure 8**. The substrate for the spherical phase singlet is a 2 inch round 0.55 mm thick D-263. The substrates in the aspheric system are 2 inch round 0.21 mm thick D-263 with broadband AR coating. The lenses themselves are written into the center of those substrates with a diameter **D** of 24.5 mm. An additional glass (1 mm thick, 50 mm diameter N-BK7 window) is utilized as a spacer for the aspheric system. Optical adhesive NOA-61 was utilized to affix the doublet components.

All three lenses were created using a direct-write laser scanner [24] and holographic replication [25, 26]. We employed the photo-alignment material LIA-CO01 (DIC Corp). All three lenses begin with a coating of a 5% solids RMM-A (Merck KGaA, $\Delta n = 0.15$ at 633 nm) in solvent propylene-glycol-methyl-ether-acetate (PGMEA). After this, the coatings of the lenses slightly differ. The singlet lens utilizes coatings of 20% solids RMM-B (Merck KGaA, $\Delta n = 0.23$ at 633 nm) in solvent PGMEA coated to achieve a half-wave retardation at 633 nm. The doublet lenses utilize coatings of 10% solids RMM-B in solvent PGMEA instead to achieve the same condition. More details associated with similar coating techniques are relayed by Xiang[20].

*Characterization of Fabricated Lenses*

After manufacture of the test lenses, the lenses were characterized individually under the microscope and with transmission. Once assembled, the spot sizes themselves are studied for both test lens structures.

Beginning first with the microscope analysis, we investigate the individual singlet spherical phase lens and aspheric lenses. The singlet reference lens, **Figure 9A**, has the expected phase pattern at the center of the lens. When we look towards the edge of the lens, **Figure 9B**, we see that the period at the edge is 2.2 μm which corresponds closely to the desired 2.16 μm.

Next, we turned to the aspheric lenses. The first aspheric lens is shown in **Figure 10A-B**. The second aspheric lens is shown in **Figure 10C-D**. Both the lenses have interesting features that traditional geometric phase patterns do not have, a small pitch region of lens 1: **Figure 10E** and phase reversal in lens 2: **Figure 10F**, which were accurately recorded into the fabricated lenses. The estimated periods at the edge of the lenses are 16.75 and 3.9 μm compared to desired values of 14.3 and 2.5 μm. In the small pitch region of lens 1, the period is approximately 2.5 μm in comparison to the desired value of 2.24 μm. Therefore, there is a variation from the desired, but the lenses themselves are holding the general shape we selected.

**Figure 10G** shows a microscope image of the assembled doublet's center. The aspheric lenses are well aligned with no noticeable offset of centers. Therefore, the phase of each individual lens should be well registered to generate the desired focusing scheme.

The transmission of individual lens components is measured broadband using white light. Each transmission is measured by selecting a region of the lens to illuminate generating a simple separation of the ±1 orders of light allowing for measurement of the 0-order of light through the element. The broadband transmission is shown in **Figure 11**. There we see a minimal response near 650 nm. At 633 nm, the transmission is 0.16% for the singlet, 2.47% for first asphere lens, and 1.77% for second asphere lens. This low zeroth order transmission indicates our lenses will perform well at the desired wavelength.

The spots of both the singlet and doublet are investigated utilizing a focal spot profiler system at 633 nm, shown in **Figure 12**. We gathered the off-axis data through rotation of the test lens on a rotation stage which is affixed using a cage mounting to a Basler ace acA380-14um camera which has 1.67 μm pixels. The simulated spots using the fabrication layout are shown in **Figure 13**. The standard spherical reference lens, **Figure 13A**, has spots ranging from diffraction limited to significant coma. The aspheric doublet, **Figure 13B**, has a larger spot on-axis, but the coma apparent in the spherical design is reduced. The measured spots are shown in **Figure 14**. The resulting ray fans appear very similar to the simulated results for the spherical singlet. The doublet appears similar for the 0° and 3° positions, before defocusing at 7°.

We created a rms calculation metric based on a weighted average of the intensity signatures at the 0°, 3°, and 7° positions. The simulated spot size radius are gathered with 30 rays and chief ray referencing to allow more fringe data into the spot size estimation, in comparison to 6 rays for the prior spot size estimates. The spot radius versus field angle position is shown in **Figure 15**. The measured asphere lens matches the trend of the simulated lens. The aspheric lens simulated rms spot size is 74.4 ± 63.2 μm and the measured rms spot size is 102 ± 36.7 μm, for just the three field angle positions. Using the three field angle position data we estimate the performance for the 9 field angle positions, the simulated result is 73.5 ± 7.06 μm and the expected measured result is 112 ± 36.9 μm. The spherical lens, **Figure 15** (blue), indicates that as we go off axis the spot size gets larger and obtains coma, but not to the extent as the simulation. This is likely because we do not have as much resolution on the actual camera, the data is digital, than the simulated results, the data is generally continuous. The simulated rms spot size is 157 ± 139 μm which compares to the measured result of 59.6 ± 36.7 μm, for just the three field angle positions. Extending to the 9 field angle positions using the 3-field angle position results, produces a simulated result of 209 ± 123 μm and an expected measured result of 74.5 ± 30.0 μm.

**DISCUSSION AND CONCLUSIONS**

In this work, we developed an optical model using MATLAB and Zemax. Beginning monochromatically, we find that the red solution from either follows the same trend, but the extension wavelengths of the Zemax do not correspond to the MATLAB results indicating the solutions found are different. To correct for the variation for extension wavelengths, we changed to adding all three wavelengths into the model. In that case we find solutions of Zemax and MATLAB models that correspond to one another. We select the Zemax polychromatic optimized solution with 1 mm spacer distance due to its reduced standard deviation of rms spot size versus the MATLAB solution. We additionally compare this lens to a standard spherical singlet of the same BFL position. The analysis of the polychromatic doublet indicates that the lens structure can be easily fabricated with high efficiency and aligned satisfactory to develop a doublet. When measured this initial prototype shows that the 7° off-axis spot is defocused in comparison to the simulated spot performance. We suspect this could be due to the waviness of the lens structures at the outside of the lenses or defects within the lenses themselves. The trend of the spot sizes follows that of the simulations where the massive coma of the spherical lens is modified by the introduction of the aspheric surfaces. Thus, this lens system affords us an entirely new aspheric focusing scheme without a master.

**LIST OF ABBREVIATIONS**

GP - geometric phase
BFL - back focal length
D - lens diameter

**ACKNOWLEGEMENTS**
We thank Dr. Shuojia Shi and Dr. Jihwan Kim for assistance in the fabrication of the test lenses and Mereck KGaA for the customized materials. This work was financially supported by ImagineOptix Corporation (NCSU grant #2014-2450). MJE owns equity in ImagineOptix.

# FIGURES AND CAPTIONS

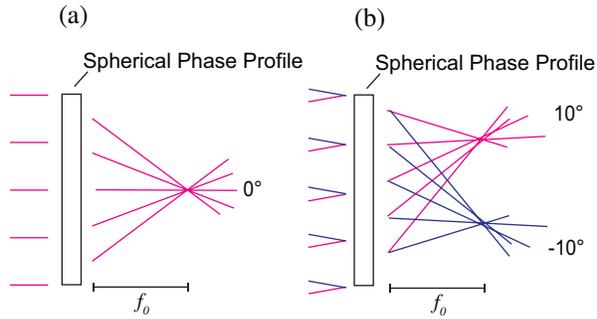

**Figure 1. Illustration of On-Axis versus Off-Axis Illumination of GP Lens Spherical Phase Profile**. Illustration of on-axis **(a)** versus off-axis (±10°) **(b)** rays through a standard spherical phase profile model of a GP lens with focal length $f_0$.

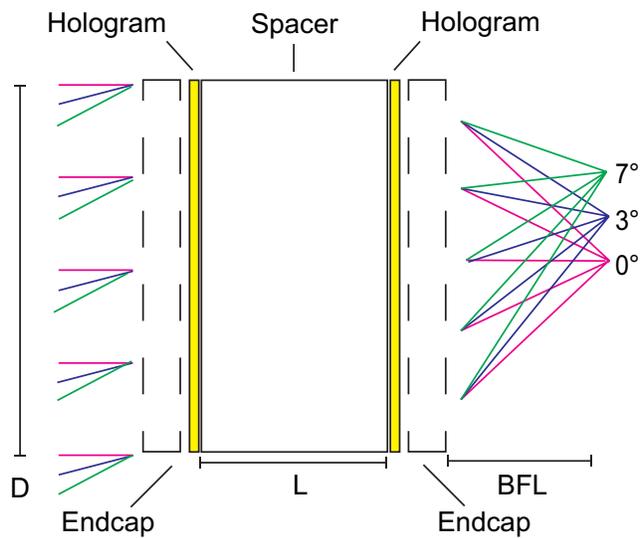

**Figure 2. Layout of the Zemax and MATLAB optimized doublet systems.** The yellow region is holographic phase pattern and liquid crystal coating which allow focusing of light for a given wavelength.

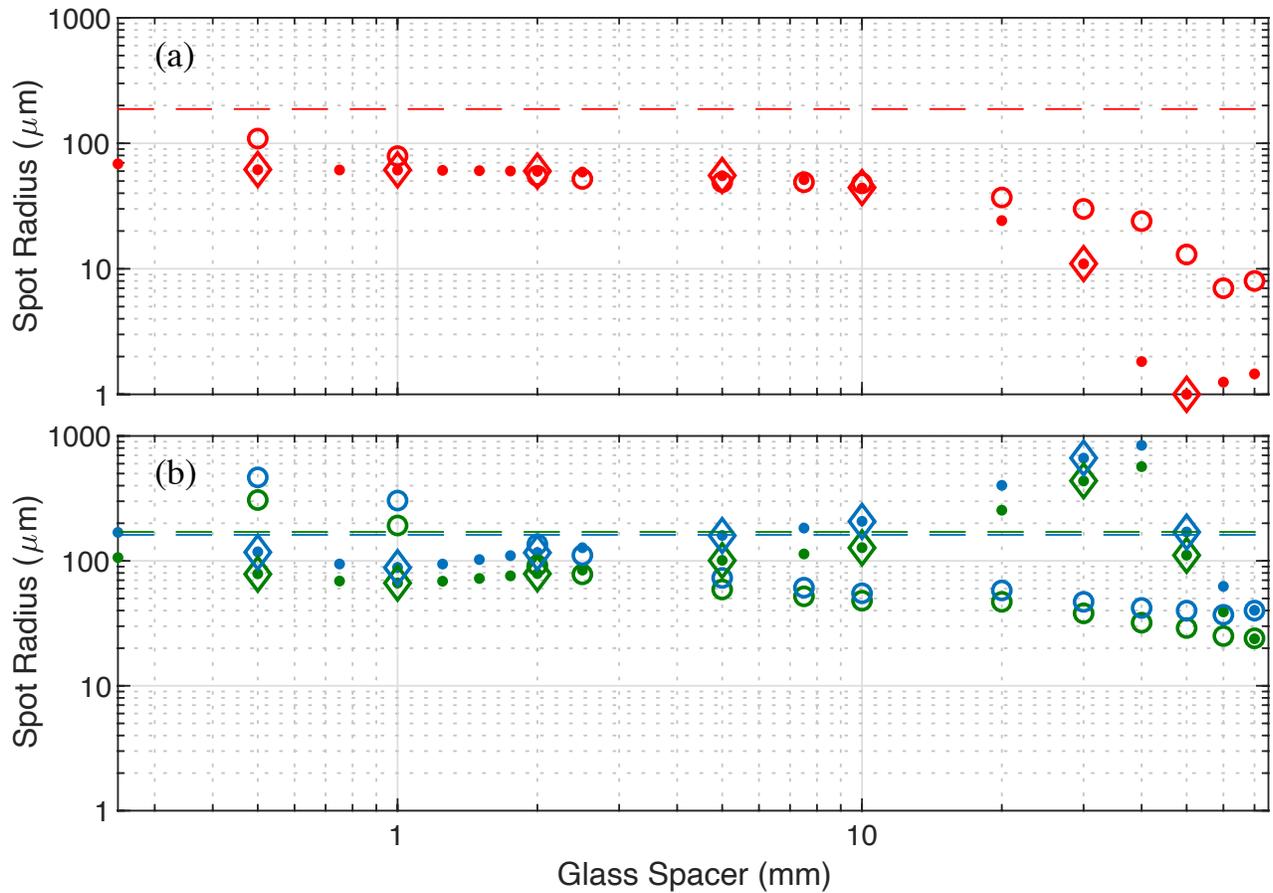

**Figure 3. *Doublet Monochromatic Aspheric Minimization comparing Zemax 16.5, Zemax 17.5, and MATLAB Models.*** Both model's doublet monochromatically minimized rms spot radius results at **(a)** 633 nm and at **(b)** 532 and 450 nm with color (red, green, and blue) indicating wavelength respectively. Dot is Zemax 16.5 generated results, diamond is Zemax 17.5 generated results, while circle is MATLAB based ray tracing code results. The averaged singlet results at 633, 532, and 450 nm are presented as dashed lines with color indicating wavelength (red, green, and blue).

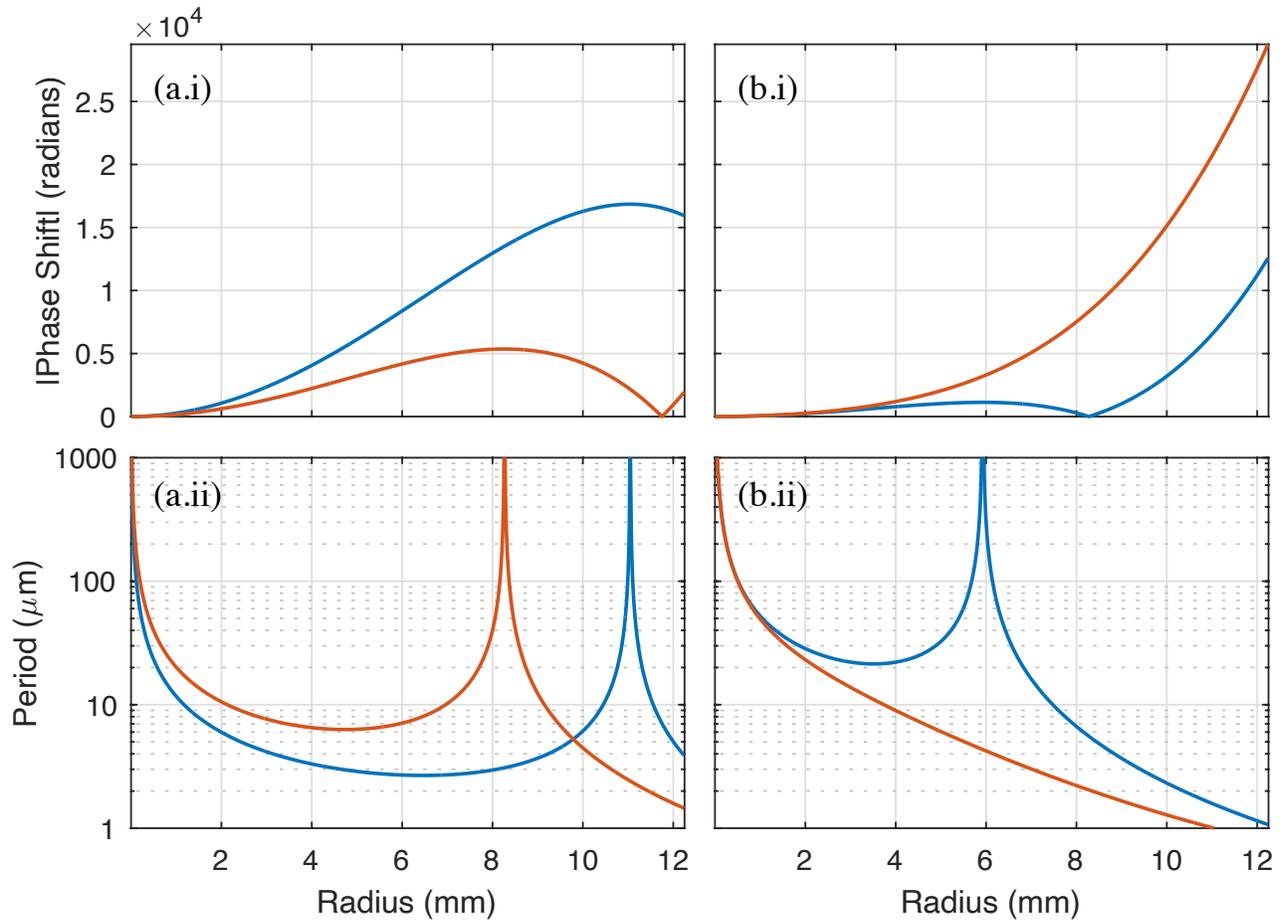

**Figure 4.** *Phase shift generated and lens period required for Monochromatic minimized Zemax 16.5 and MATLAB Models.* Resulting absolute value of phase shift **(.i)** and period **(.ii)** resulting from minimized Zemax 16.5 **(a)** and MATLAB **(b)** models for a monochromatic doublet with 1 mm separation. The profiles of lens 1 shown in blue, while the lens 2 profiles are shown in red.

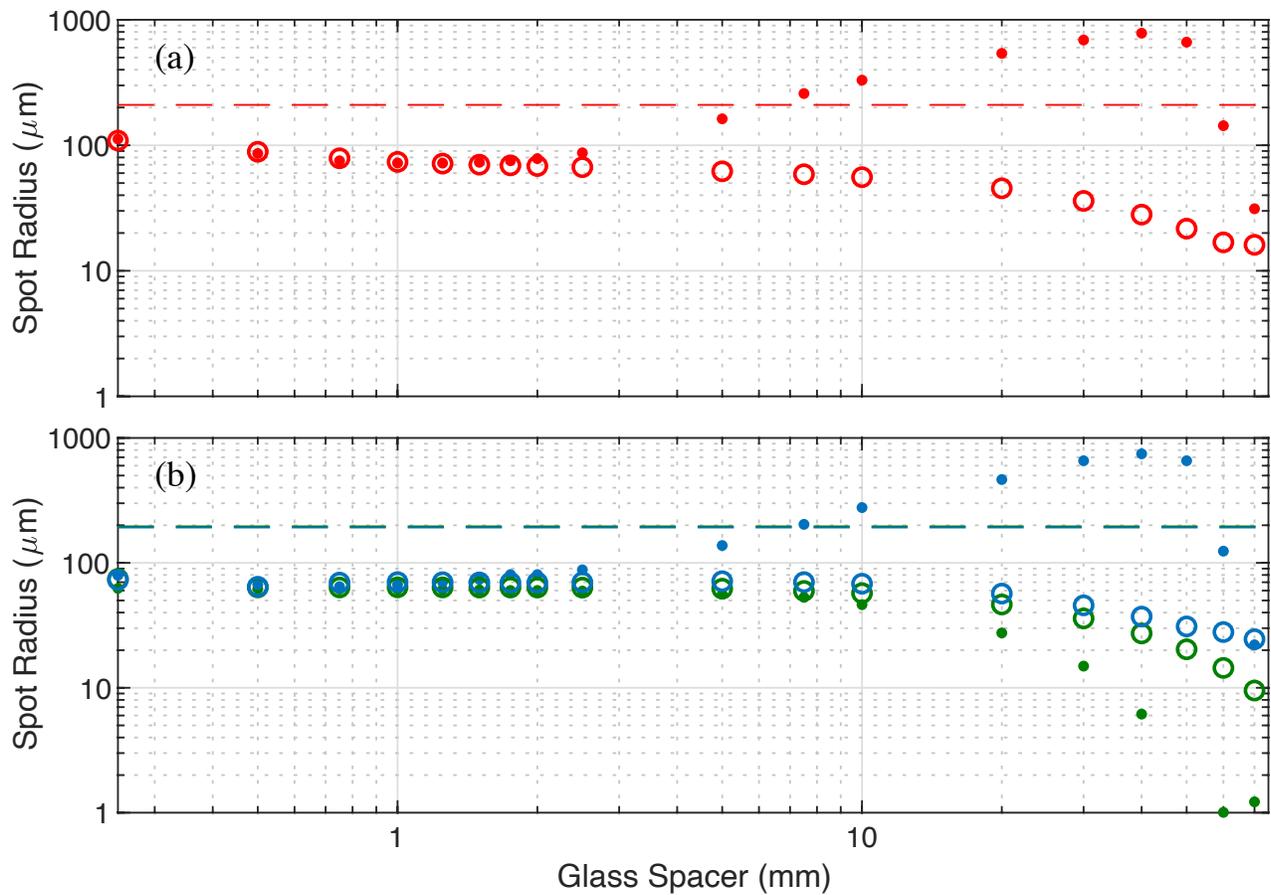

*Figure 5. Doublet Polychromatic Aspheric Minimization for Both Zemax versions.* Both versions of Zemax doublet polychromatically minimized rms spot radius results at **(a)** 633 nm and **(b)** 532 and 450 nm with color (red, green, and blue) indicating wavelength respectively. Dot is Zemax 16.5 generated results while circle is Zemax 17.5 results. The averaged singlet result at 633, 532, and 450 nm is presented as dashed lines with color indicating wavelength (red, green, and blue).

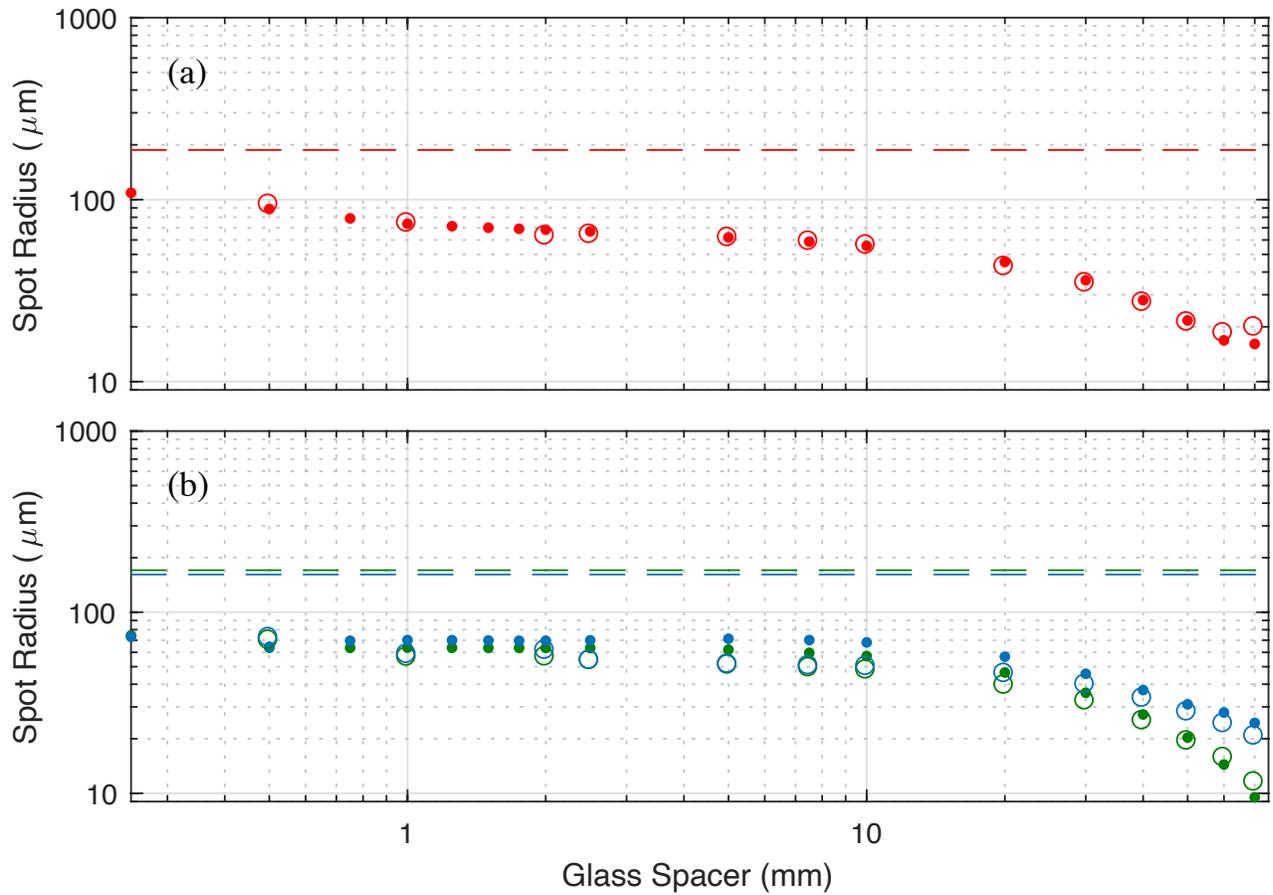

**Figure 6.** *Doublet Polychromatic Aspheric Minimization comparing Zemax 17.5 and MATLAB Models.* Both model's doublet polychromatically minimized rms spot radius results at **(a)** 633 nm and at **(b)** 532 and 450 nm with color (red, green, and blue) indicating wavelength respectively. Dot is Zemax 17.5 generated results while circle is MATLAB based ray tracing code results. The averaged singlet result at 633, 532, and 450 nm is presented as dashed lines with color indicating wavelength (red, green, and blue).

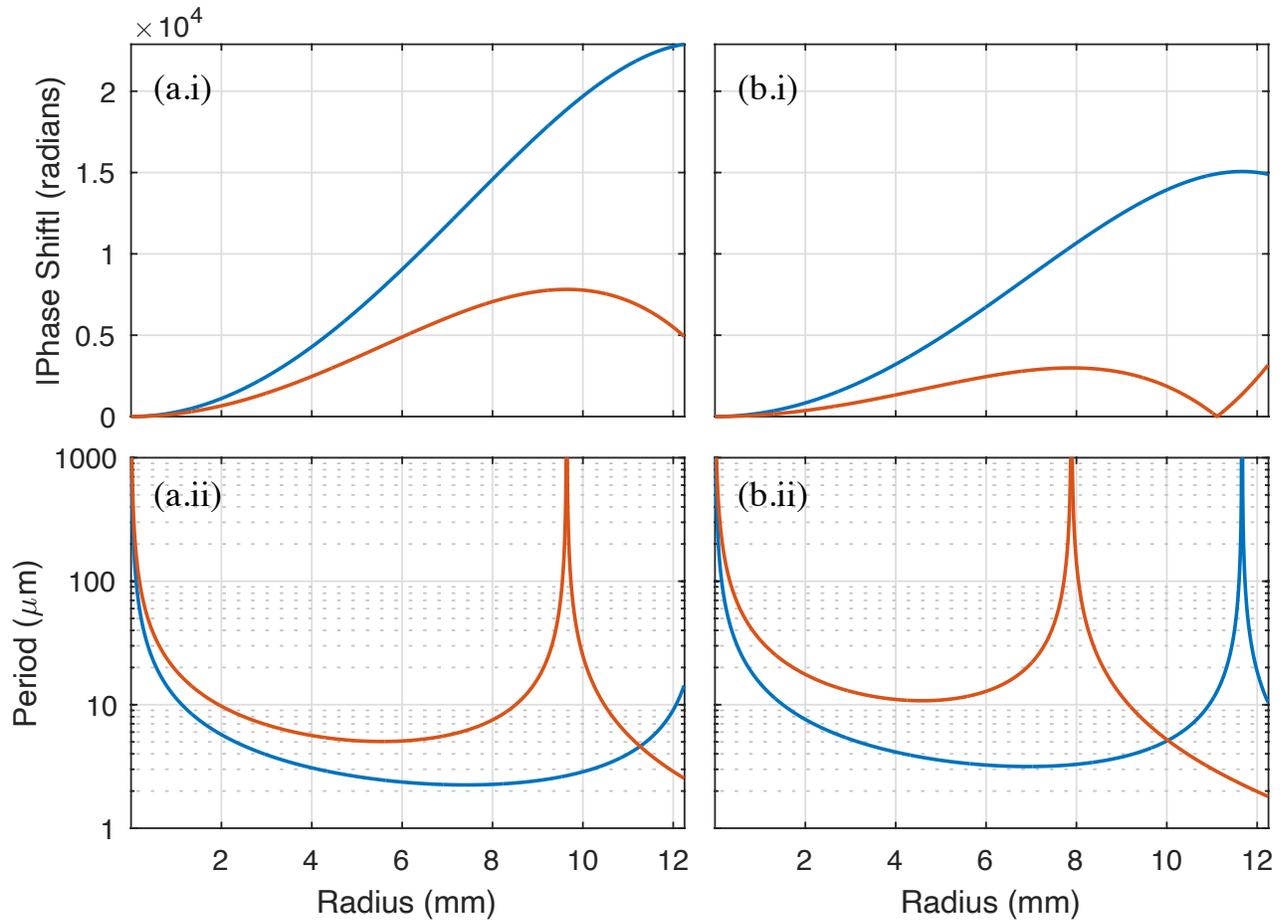

*Figure 7. Phase shift generated and lens period required for Polychromatic Minimized Zemax 17.5 and MATLAB Models.* Resulting absolute value of phase shift (**.i**) and period (**.ii**) resulting from minimized Zemax 17.5 **(a)** and MATLAB **(b)** models for a polychromatic doublet with 1 mm separation. The profiles of lens 1 shown in blue, while the lens 2 profiles are shown in red.

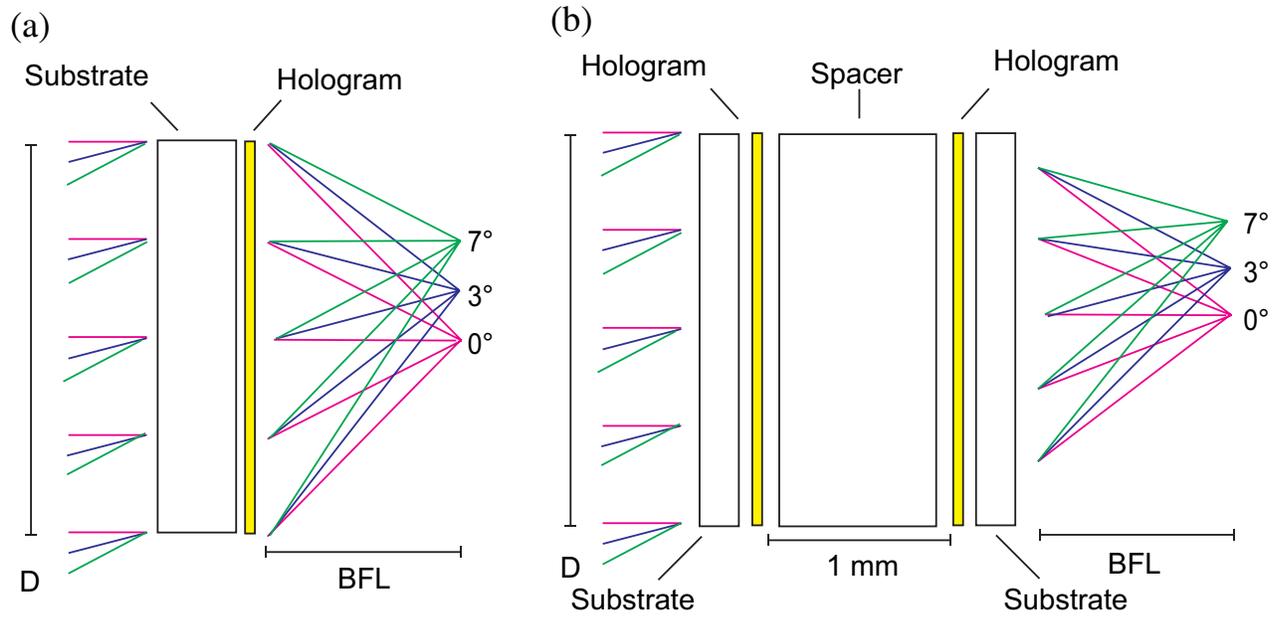

**Figure 8.** *Aspheric Lenses Fabrication Layout.* The layout of fabricated lenses with **(a)** the singlet reference and **(b)** the doublet system. The yellow region is holographic phase pattern and liquid crystal coating which allow focusing of light for a given wavelength.

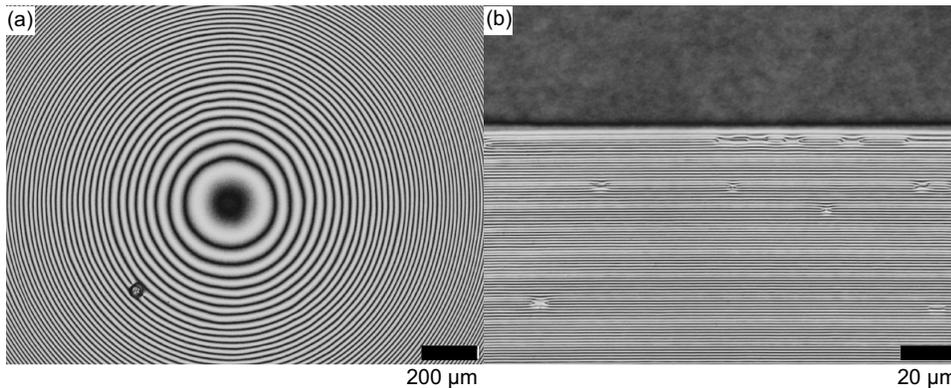

**Figure 9.** *Singlet Reference Microscope Images.* Microscope images of the **(a)** center and **(b)** edge of the singlet reference lens.

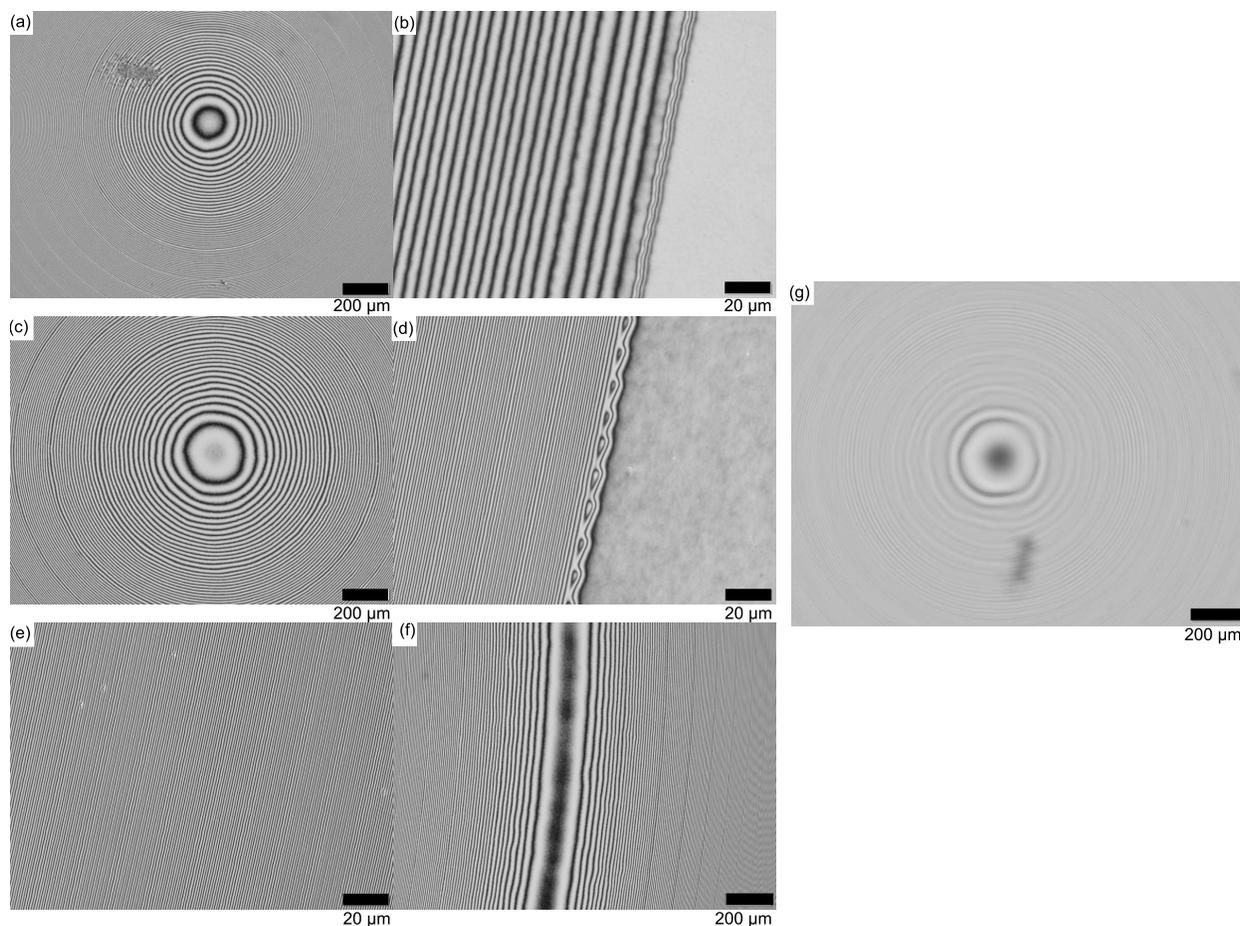

**Figure 10.** *Doublet Microscope Images.* Microscope images of the **(a)** center and **(b)** edge of lens 1 of the doublet. Microscope images of the **(c)** center and **(d)** edge of lens 2 of the doublet. Images of detailing distinct features within each fabricated aspheric lens with **(e)** the small pitch region of lens 1 and the **(f)** phase reversal region of lens 2 of the doublet. Microscope image of central alignment of doublet **(g)**.

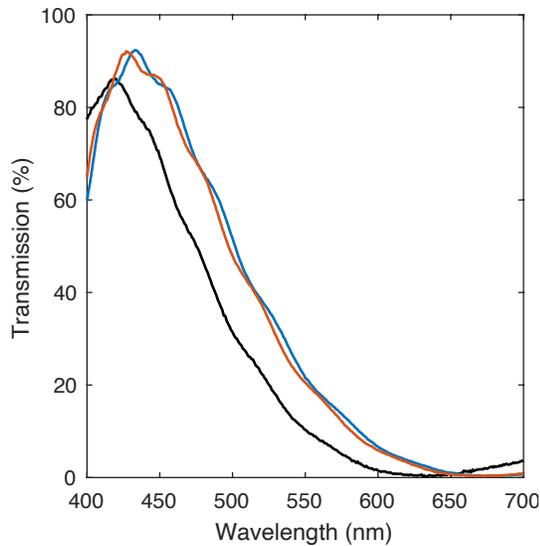

**Figure 11.** *Zeroth order transmission of individual lens components*. Zeroth order transmission of individual lens components prior to assembly where black is the standard spherical singlet while blue and red correspond to lens 1 and 2 of the doublet respectively.

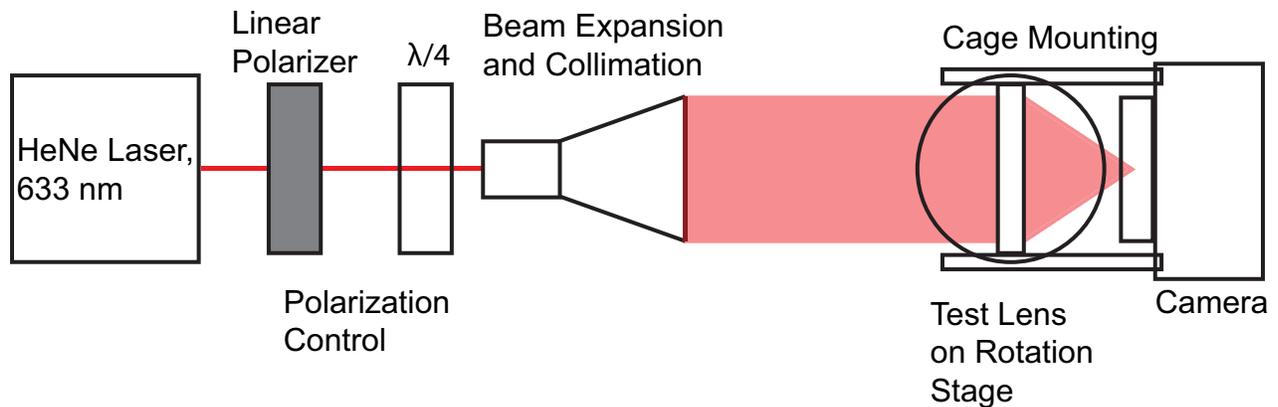

**Figure 12.** *Beam Profiler Test System.* Layout of the focal spot profiler for imaging the focus spots of the lens.

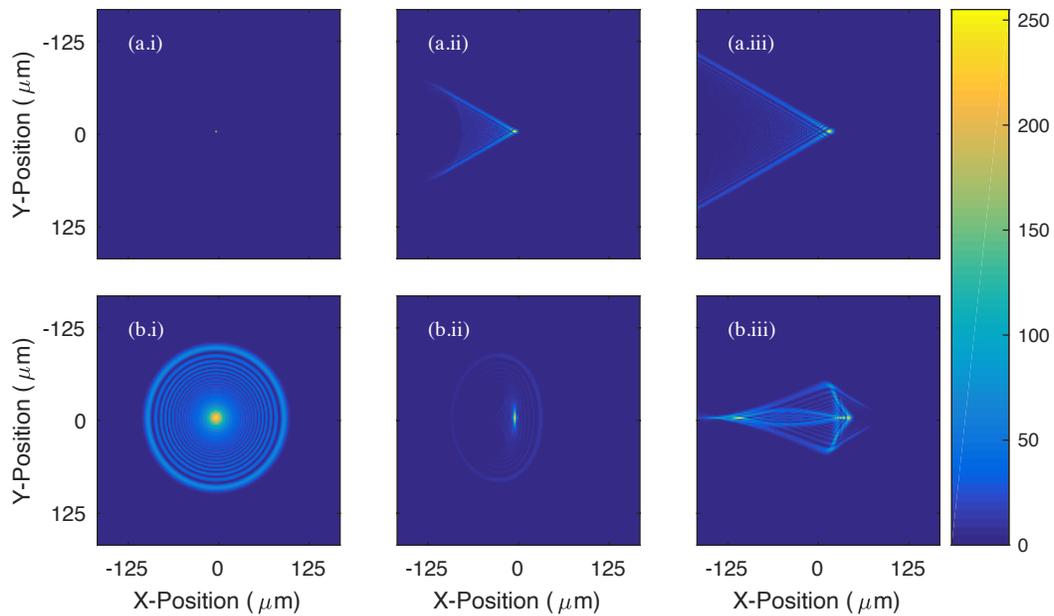

**Figure 13.** *Simulated Lens Focused Spots.* Simulations of the **(a)** standard spherical singlet reference and **(b)** aspheric doublet as assembled showing the **(.i)** on-axis, **(.ii)** 3° off-axis spots and **(.iii)** 7° off-axis spot.

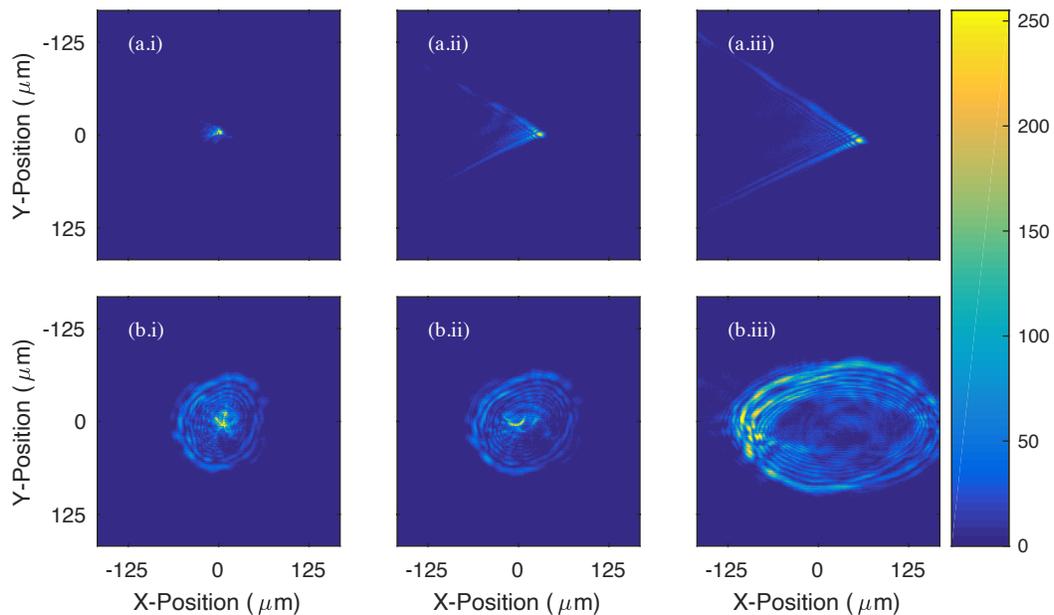

**Figure 14.** *Measured Lens Focused Spots.* Spot performance measurements using the focal spot profiler of the **(a)** standard spherical singlet reference and **(b)** aspheric doublet showing the **(.i)** on-axis, **(.ii)** 3° off-axis spots and **(.iii)** 7° off-axis spots.

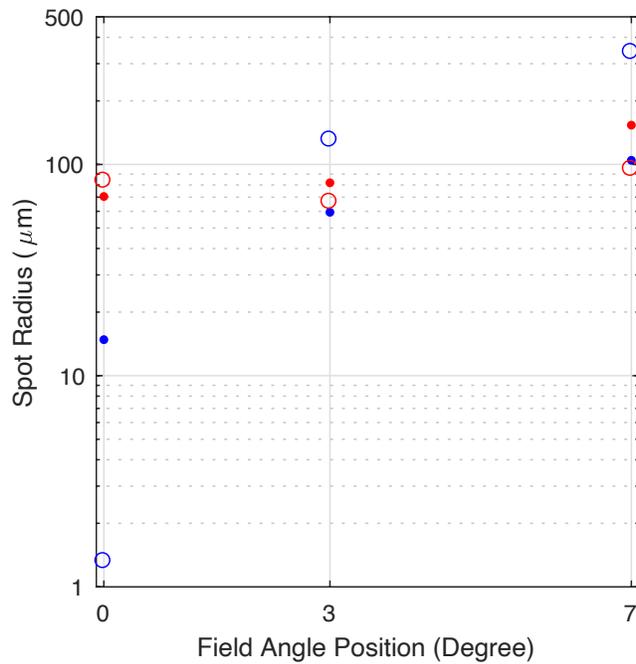

**Figure 15.** *Comparison of measured versus simulated rms spot radius.* RMS spot radius at 633 nm across different search mechanisms utilized in this work for 0°, 3°, and 7° field angle positions. Blue is spherical reference model while red is aspheric polynomial doublet results. Circle is Zemax simulation results while dot is measured results.